\documentstyle[twocolumn, epsfig, floatfig, array, floats, aps]{revtex}

\begin{document}

\newcommand{\be}{${}^9{\rm Be}^+\:$}
\newcommand{\uk}{|\!\!\uparrow\rangle}
\newcommand{\ub}{\langle\uparrow\!\!|}
\newcommand{\dk}{|\!\!\downarrow\rangle}
\newcommand{\db}{\langle\downarrow\!\!|}
\newcommand{\uuk}{|\!\!\uparrow\uparrow\rangle}
\newcommand{\uub}{\langle\uparrow\uparrow\!\!|}
\newcommand{\ddk}{|\!\!\downarrow\downarrow\rangle}
\newcommand{\ddb}{\langle\downarrow\downarrow\!\!|}
\newcommand{\udk}{|\!\!\uparrow\downarrow\rangle}
\newcommand{\udb}{\langle\uparrow\downarrow\!\!|}
\newcommand{\duk}{|\!\!\downarrow\uparrow\rangle}
\newcommand{\dub}{\langle\downarrow\uparrow\!\!|}
\newcommand{\ua}{\uparrow}
\newcommand{\da}{\downarrow}

\newcommand{\init}{|\psi_i\rangle}
\newcommand{\dfs}{|\psi_{\rm DFS}\rangle}
\newcommand{\psip}{|\psi_+\rangle}
\newcommand{\psin}{|\psi_-\rangle}

\newcommand{\jprl}{\emph{Phys. Rev. Lett. }}
\newcommand{\jpra}{\emph{Phys. Rev. A }}

\title{Recent Results in Trapped-Ion Quantum Computing at NIST\\}

\author{D. Kielpinski, A. Ben-Kish, J. Britton, V. Meyer, M.A. Rowe, C.A.
Sackett${}^1$, W.M. Itano, C. Monroe${}^2$, and D.J. Wineland\\}

\address{Time and Frequency Division, National Institute of Standards and Technology,
Mailstop 847, 325 Broadway, Boulder CO 80303, USA\\E-mail:
davidk@boulder.nist.gov\\
${}^1$ Department of Physics, University of Virginia, Charlottesville VA 22904,
USA\\
${}^2$ Department of Physics, University of Michigan, Ann Arbor MI 48109, USA\\} 

\maketitle

\begin{abstract}

We review recent experiments on entanglement, Bell's inequality, and
decoherence-free subspaces in a quantum register of trapped \be ions. We have
demonstrated entanglement of up to four ions \cite{cass} using the technique of
M{\o}lmer and S{\o}rensen \cite{molmer}. This method produces the state
$(\uuk+\ddk)/\sqrt{2}$ for two ions and the state $(|\! \da\da\da\da \rangle +
\, | \! \ua\ua\ua\ua \rangle)/\sqrt{2}$ for four ions. We generate the entanglement
deterministically in each shot of the experiment. Measurements on the two-ion
entangled state violates Bell's inequality at the $8\sigma$ level \cite{mary}.
Because of the high detector efficiency of our apparatus, this experiment closes
the detector loophole for Bell's inequality measurements for the first time.
This measurement is also the first violation of Bell's inequality by massive
particles that does not implicitly assume results from quantum mechanics.
Finally, we have demonstrated reversible encoding of an arbitrary qubit,
originally contained in one ion, into a decoherence-free subspace (DFS) of two
ions \cite{dfs}. The DFS-encoded qubit resists applied collective dephasing
noise and retains coherence under ambient conditions 3.6 times longer than does
an unencoded qubit. The encoding method, which uses single-ion gates and the
two-ion entangling gate, demonstrates all the elements required for two-qubit
universal quantum logic.\\ \\

\end{abstract}

\section{Introduction}

Trapped ions are a promising candidate system for implementing quantum
computing (QC), with potential applications like efficient factorization
of large numbers \cite{shor} and efficient searching of large databases
\cite{grover}. While large-scale QC is still far off, trapped-ion quantum
registers have already demonstrated all the essential ingredients of QC, including
initialization to a known quantum state \cite{diedrich,brian,cacool}, efficient
detection of final states \cite{mary,cacool}, long qubit coherence times
\cite{dfs,projection,fisk,cacohere}, and a universal set of quantum logic gates
\cite{cass,cacool,xor,address}. Recent work on trapped-ion QC at NIST has used
this basic set of tools to produce four-particle entanglement \cite{cass},
observe violation of Bell's inequality \cite{mary}, and encode quantum
information into a decoherence-free subspace \cite{dfs}.\\ \\

In the experiments described below, the qubits are electronic states of \be ions
held in a linear RF trap \cite{linear} of the type described in reference
\cite{quentin}. The relevant electronic energy levels of the \be ion are shown
in Fig. 1. Each ion encodes one qubit in its ${}^2{\rm S}_{1/2}|F=2,
m_F=-2\rangle$ and ${}^2{\rm S}_{1/2}|F=1, m_F=-1\rangle$ hyperfine sublevels,
denoted by $\dk$ and $\uk$, respectively. The ions are readily initialized in
$\dk$ by optical pumping. We detect the number of ions in state $\dk$ by
applying left circularly polarized laser light to excite the $\dk$ - ${}^2{\rm
P}_{3/2}$ transition at about 313 nm. The polarization of the light ensures
excitation to the ${}^2{\rm P}_{3/2}|F=3, m_F=-3\rangle$ hyperfine sublevel,
which can only decay to the $\dk$ sublevel by dipole selection rules. Thus an
ion in state $\dk$ can scatter thousands of photons without changing its
hyperfine state. On the other hand, state $\uk$ is spectrally resolved from
$\dk$ by the hyperfine splitting of 1.25 GHz and therefore scatters very little
light. The resulting detection accuracy per ion is about 98\% in a single shot
of the experiment.\\ \\

\begin{figure}

\begin{center}

\epsfig{file=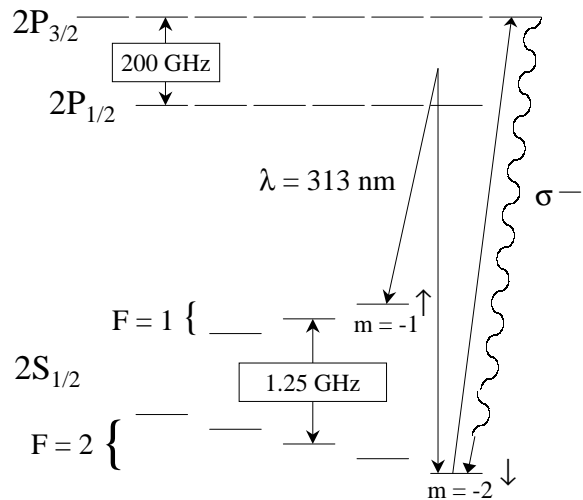, width=8cm, clip=, bbllx=60, bblly=180,
bburx=550, bbury=620}

\caption{Partial level diagram of the \be ion, showing the cycling transition for $\sigma_-$
resonant light and the two-photon off-resonant coupling for Raman transitions. Wavelengths
for all transitions are approximately 313 nm.}

\end{center}

\end{figure}

We produce controlled rotations of a single qubit by driving stimulated Raman
transitions between states $\uk$ and $\dk$. The laser beams driving the transitions
are tuned about midway between the ${}^2{\rm P}_{3/2}$ and ${}^2{\rm P}_{1/2}$
states. If the frequency difference $\nu_{\rm diff}$ between the Raman laser beams
is equal to the energy splitting $\nu_0$ between states $\uk$ and $\dk$, we
observe Rabi flopping between $\uk$ and $\dk$ corresponding to the evolution

\begin{eqnarray}
\dk &\rightarrow& \cos\frac{\theta}{2}\,\dk +
e^{i\phi}\sin\frac{\theta}{2}\,\uk \label{careq}\\
\uk &\rightarrow& \cos\frac{\theta}{2}\,\uk -
e^{-i\phi}\sin\frac{\theta}{2}\,\dk \nonumber
\end{eqnarray}

\noindent Here $\theta$ is proportional to the Raman pulse duration and $\phi$
is the phase difference between the Raman beams at the position of the ion,
referred to as the ``ion phase". For two ions, we write the ion phases as
$\phi_1, \phi_2$. We can effect common-mode changes of $\phi_1$ and $\phi_2$ by
changing the phase of the RF synthesizer that controls the Raman difference
frequency. By changing the strength of the trap, we can control the ion spacing
precisely, allowing us to make differential changes to $\phi_1$ and $\phi_2$.
These two techniques combined allow us to control $\phi_1$ and $\phi_2$
independently for two ions, so that we can perform single-qubit rotations on one
ion without affecting the other \cite{dfs}.\\ \\

The potential created by the ion trap is harmonic with axial frequency $\nu_{\rm
CM}$ of about 5 MHz. Doppler cooling allows us to cool the ions to temperatures
$T \sim h \nu_{\rm CM}/k_B$. The Doppler-cooled ions are strongly coupled by the
Coulomb interaction and form a rigid crystal, which in these experiments is a
string of ions lying along the trap axis. Because the ions are strongly coupled,
their quantized motion is best described in terms of normal modes with
frequencies $\nu_i$. For two ions, for instance, the motion of the ions along
the axis decomposes into symmetric and antisymmetric normal modes at frequencies
$\nu_{\rm CM}$ and $\sqrt{3}\nu_{\rm CM}$. The lowest-lying normal mode for any
number of ions is the center-of-mass mode at frequency $\nu_{\rm CM}$. In this
mode the ion string moves as a unit.\\ \\

We can control the motional quantum state of the ions by driving stimulated
Raman transitions that couple spin and motion. In our geometry, the wavevector
difference between Raman beams lies along the axis of the ion trap, so the Raman
beams exert a dipole force only along the axis. Setting $\nu_{\rm diff}$ equal
to $\nu_0 + m \nu_i$ ($m$ an integer) induces Rabi flopping between states
$\dk|n\rangle$ and $\uk|n+m\rangle$ with evolution like that of
Eq.~(\ref{careq}). Here $|n\rangle$ denotes the $n$th Fock state of the $i$th
normal mode. In particular, we can cool the ions by driving a pulse with
$\theta=\pi$ at $\nu_{\rm diff} = \nu_0 - \nu_i$ (the ``red sideband") and
optically pumping from $\uk$ back to $\dk$. Repeating the cycle several times
permits preparation of the ion string in its motional ground state
\cite{diedrich,brian}.\\ \\

\section{Generating Entanglement}

We have entangled strings of up to four ions \cite{cass} using the technique
proposed by S{\o}rensen and M{\o}lmer \cite{molmer}. In this method, one drives
two Raman transitions simultaneously, with the same Rabi frequency on all ions.
Fig. 2 shows an energy-level diagram including the relevant transitions. The
detuning $\delta$ ensures that population transfer between motional states is
nonresonant and therefore small. However, the sum of the two Raman transition
frequencies is $2\nu_{\ua\da}$, so that for any pair of ions the process $\ddk
\leftrightarrow \uuk$ is resonant. For $N$ ions initially in $\dk$, the
resulting evolution produces the state\\

\begin{equation}
\begin{array}{llll}
\label{molstate}
|\psi_e\rangle &=& (\ddk+\uuk)/\sqrt{2} & \quad \mbox{(two\ ions)} \\
& & (|\!\da\da\da\da\rangle +\, |\!\ua\ua\ua\ua\rangle)/\sqrt{2} & \quad \mbox{(four\ ions)}
\end{array}
\end{equation}

\noindent both of which are maximally entangled states.\\ \\

\begin{figure}

\begin{center}

\epsfig{file=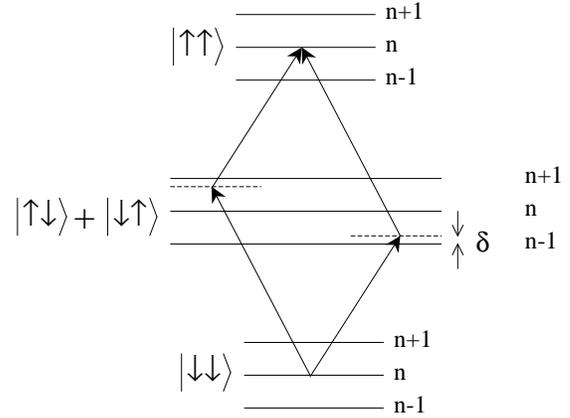, width=8cm, clip=, bbllx=70, bblly=230,
bburx=510, bbury=570}

\caption{Level diagram for the S{\o}rensen-M{\o}lmer entangling gate in the case
of two ions. Here, as opposed to Fig. 1, arrows indicate two-photon Raman
transitions. Raman transitions near both red and blue sidebands are applied
simultaneously to create the state $(|\da\da\rangle + |\ua\ua\rangle)/\sqrt{2}$.}

\end{center}

\end{figure}

To implement this technique, we pass one of the Raman beams through an
electro-optic modulator (EOM) to generate the two required Raman difference
frequencies and uniformly illuminate the ion string with both Raman beams. For
two ions, we tune the Raman difference frequencies close to the motional
sidebands of the antisymmetric mode; for four ions, we use the mode with
frequency $\sqrt{29/5} \; \nu_{\rm CM}$. In both cases, the mode we choose has
equal amplitudes of motion for all ions, so that the Raman Rabi frequencies for
all ions are the same under uniform illumination.\\ \\

In general, the state produced by this method is characterized by some density
matrix $\rho$ that approximates the desired state $|\psi_e\rangle$. To show
entanglement of the experimental state $\rho$, it is sufficient to show that the
fidelity $F = \langle\psi_e|\rho|\psi_e\rangle$ exceeds $1/2$ \cite{cass,dur}. We have

\begin{equation}
\label{fideq}
F = (\rho_{\da\da\ldots\da\da} + \rho_{\ua\ua\ldots\ua\ua})/2 +
|\rho_{\da\da\ldots\ua\ua}|
\end{equation}

\noindent We measure the diagonal density matrix elements
$\rho_{\da\ldots\da,\da\ldots\da}, \rho_{\ua\ldots\ua,\ua\ldots\ua}$ by detecting the number
of ions in $\dk$. To find the far-off-diagonal coherence
$|\rho_{\da\ldots\da\ua\ldots\ua}|$, we rotate all ions together as in Eq.~(\ref{careq})
with $\theta=\pi/2$ and with phases $\phi_i = \phi$ relative to the entangling
pulse. Then we measure the parity $\Pi$ of the ion string, defined as $+1$ for
an even number of ions in $\dk$, $-1$ for an odd number in $\dk$. It is easy to show that

\begin{equation}
\label{pareq}
\Pi = 2 |\rho_{\da\da\ldots\ua\ua}| \cos N\phi
\end{equation}

\noindent for $N$ ions, if all other off-diagonal coherences are absent.
Sweeping the phase of the RF synthesizer that generates the Raman difference
frequency gives a sinusoidal signal of $\Pi$ vs. $\phi$, as shown in Fig. 3. We
extract the coherence $|\rho_{\da\da\ldots\ua\ua}|$ by fitting a sine wave to
the data of Fig. 3. The density matrix elements needed to calculate the fidelity
are given in Table 1 for two and four ions. For two ions we find $F =
0.83\pm0.01$, while for four ions $F = 0.57\pm0.02$, demonstrating entanglement
in both cases \cite{cass}. It is important to note that the data used to
calculate $F$ are not post-selected. Since the evolution is deterministic, our
method produces an entangled state in each shot of the experiment.\\ \\

\begin{table}[h]

\caption{Partial density matrix of the state produced by the entangling pulse.
From Eq.~(\ref{fideq}), only these matrix elements are needed to determine
whether the experimentally generated state is entangled. Errors are $\pm 0.01$
for two ions and $\pm 0.02$ for four ions.}

\vspace{0.5cm}

\begin{center}

\begin{tabular}{|cccc|}
\hline
Number of Ions & $\rho_{\da\da\ldots\da\da}$ & $\rho_{\ua\ua\ldots\ua\ua}$
& $|\rho_{\da\da\ldots\ua\ua}|$\\
2 & 0.46 & 0.385 & 0.43\\
4 & 0.35 & 0.35 & 0.215\\
\hline
\end{tabular}

\end{center}

\end{table}

\begin{figure}[top]

\begin{center}

\epsfig{file=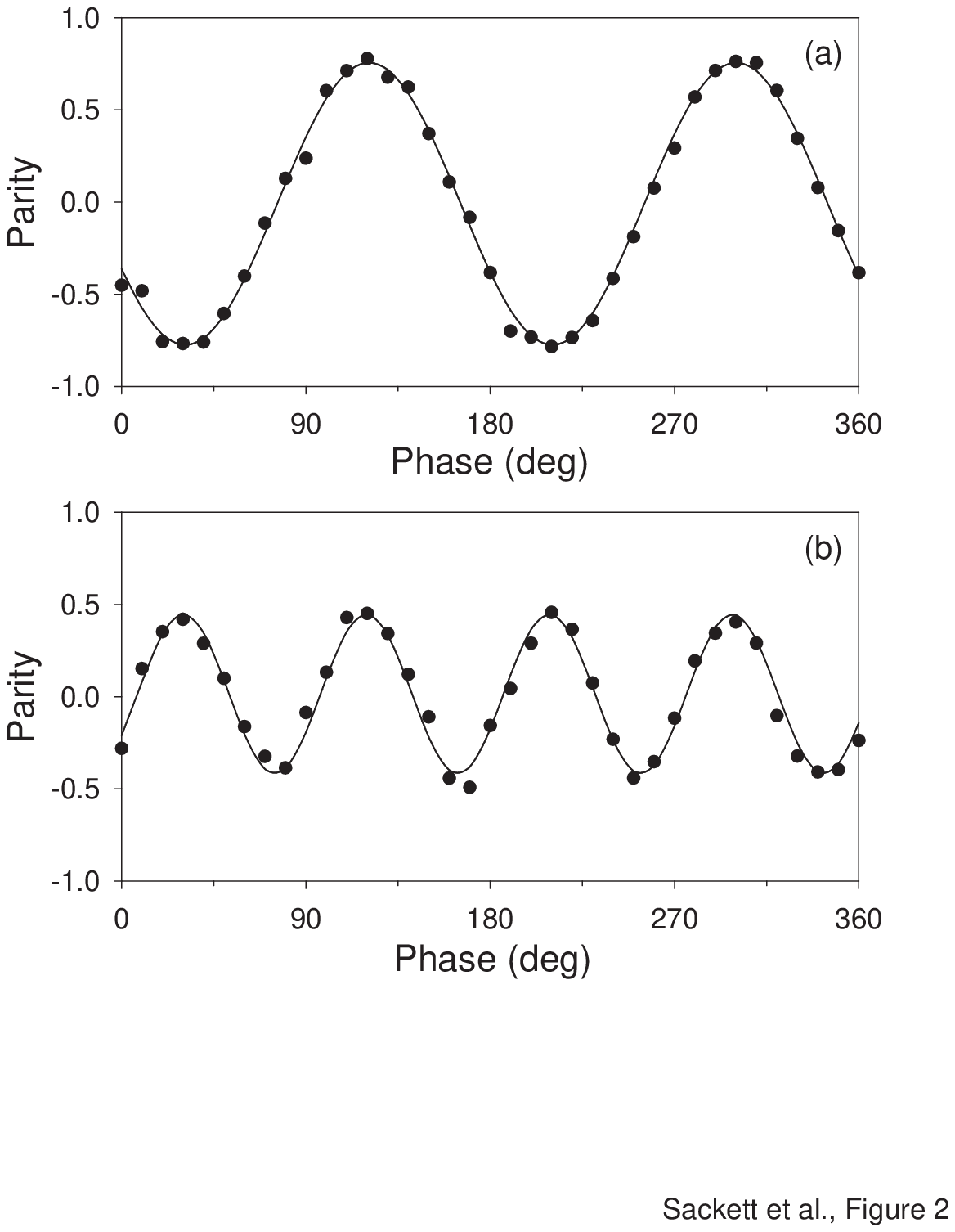, width=8cm, clip=, bbllx=30, bblly=130,
bburx=370,bbury=500}

\caption{Measurements of the parity $\Pi$ as a function of RF synthesizer phase
$\phi$ for two and four ions. The data show clear sinusoidal dependence as in
Eq.~(\ref{pareq}), allowing us to extract the far-off-diagonal coherence
$|\rho_{\da\da\ldots\ua\ua}|$.}

\end{center}

\end{figure}

\section{Violation of a Bell Inequality}

Classical physics obeys the principle of local realism, which states that
objects have definite properties whether or not they are measured and that
measurements of an object's properties are not affected by events taking place
sufficiently far away. As shown by Einstein, Podolsky, and Rosen \cite{epr},
quantum mechanics is incomplete if we assume local realism. Bell \cite{bell}
proved that quantum mechanics is actually inconsistent with local realism.
Quantitative conditions for falsifying local realism by experimental
measurements were first proposed by Bell \cite{bell} and were developed further
by Clauser, Horne, Shimony, and Holt (CHSH) \cite{chsh}. In such an experiment,
one prepares a pair of particles in a repeatable way, subjects them to
independent classical manipulations parametrized by variables $\Phi_1, \Phi_2$,
and measures a classical property yielding a value $P_1(\Phi_1), P_2(\Phi_2) =
\pm 1$ for each particle. Defining the correlation function $q(\Phi_1,\Phi_2) =
P_1(\Phi_1)P_2(\Phi_2)$, one finds that the Bell inequality

\begin{eqnarray}
\label{belleq}
B(\alpha_1,\delta_1,\beta_2,\gamma_2) &=& |q(\delta_1,\gamma_2) -
q(\alpha_1,\gamma_2)|\\
&& + |q(\delta_1,\beta_2) + q(\alpha_1,\beta_2)| \leq 2 \nonumber
\end{eqnarray}

\noindent holds for any values $\alpha_1,\delta_1$ (resp. $\beta_2,\gamma_2$)
of $\Phi_1$ (resp. $\Phi_2$) under the assumption of local realism.\\ \\

In our experiment \cite{mary}, we apply the entangling operation of
Eq.~(\ref{molstate}) to two ions to prepare our initial state. Then we perform a
carrier rotation as in Eq.~(\ref{careq}) with $\theta=\pi/2$ and phases
$\phi_1$, $\phi_2$ relative to the entangling pulse phases. Finally, we detect
the number of bright ions. The
calibration of $\phi_1$ and $\phi_2$ against the trap strength and the RF
synthesizer phase were performed using assumptions from classical physics only
\cite{mary}. We can therefore identify $\phi_1,\phi_2$ with the classical
control parameters $\Phi_1,\Phi_2$, so the experiment is of the type proposed by
Bell and CHSH. Quantum mechanics predicts that, for an ideal realization of our experiment,
$B(\alpha_1,\delta_1,\beta_2,\gamma_2)$ attains a maximum value of $2\sqrt{2}$
for

\begin{equation}
\label{magic}
\begin{array}{lllllll}
\alpha_1 &=& -\pi/8 & \qquad & \delta_1 &=& 3\pi/8 \\
\beta_2 &=& -\pi/8 & \qquad & \gamma_2 &=& 3\pi/8
\end{array}
\end{equation}

\noindent contradicting the bound $B(\alpha_1,\delta_1,\beta_2,\gamma_2) \leq
2$ required by local realism.\\ \\

We repeated the experiment $N_{\rm tot} =$ 20 000 times for each of the four
sets of phases given by Eq.~(\ref{magic}). These data are presented as
histograms in Fig. 4. We see that the three cases of zero, one, and two ions
bright are readily distinguishable. The overlap of histograms corresponding to
the three cases gives an inaccuracy of 2\% in discriminating between the three
cases. This inaccuracy mostly arises from pumping of the $\dk$ state to the
$\uk$ state because of imperfect polarization of the detection light. Counting
the number of events $N_0$ ($N_1,N_2$) with zero (one, two) ions bright, we
extract the correlation function using

\begin{equation}
\label{corr}
q = \frac{(N_0+N_2)-N_1}{N_{\rm tot}}
\end{equation}

\noindent and obtain a set of four values of $q(\phi_1,\phi_2)$. We repeated the
entire procedure five times to ensure that the measured Bell signal did not vary
significantly from run to run. The five sets of correlation functions and the resulting Bell
signals are shown in Table 2. The statistical error of each Bell signal was $\pm
0.01$. Averaging the Bell signals together, we find

\begin{equation}
\label{violation}
B(-\frac{\pi}{8}, \frac{3\pi}{8}, -\frac{\pi}{8}, \frac{3\pi}{8}) = 2.25\pm0.03
\end{equation}

\noindent an $8\sigma$ violation of the Bell inequality Eq.~(\ref{belleq}).
The error in the average Bell signal significantly exceeds the error
of each individual measurement of the Bell signal. This discrepancy arises from
errors in setting the phases of Eq.~(\ref{magic}).\\ \\

\begin{figure}[top]

\begin{center}

\epsfig{file=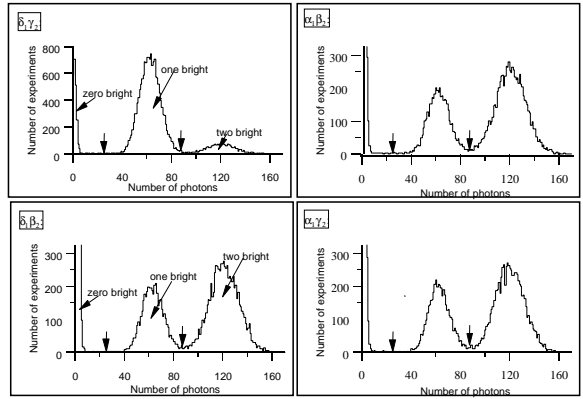, width=8cm, clip=, bbllx=100, bblly=330,
bburx=500, bbury=620}

\caption{Histograms of fluorescence signals for the four sets of phases given in
Eq.~(\ref{magic}). The vertical arrows break up the data into cases of zero,
one, or two ions bright with 98\% accuracy.}

\end{center}

\end{figure}

\begin{table}[h]

\begin{center}

\caption{The correlation values and resulting Bell's signals for five runs of
the experiment. The phase angles $\alpha_{1}, \delta_{1}, \beta_{2}, \gamma_{2}$
are given by Eq.~(\ref{magic}). The statistical errors are 0.006 and 0.012 for
the values of $q$ and $B$ respectively.}

\vspace{0.5cm}

\begin{tabular}{|ccccc|}
\hline
$q(\alpha_{1},\beta_{2})$ &$q(\alpha_{1},\gamma_{2})$
&$q(\delta_{1},\beta_{2})$
&$q(\delta_{1},\gamma_{2})$ &$B(\alpha_{1},\delta_{1},\beta_{2},\gamma_{2})$ \\
0.541 &0.539 &0.569 &-0.573 &2.222 \\
0.575 &0.570 &0.530 &-0.600 &2.275 \\
0.551 &0.634 &0.590 &-0.487 &2.262 \\
0.575 &0.561 &0.559 &-0.551 &2.246 \\
0.541 &0.596 &0.537 &-0.571 &2.245 \\
\hline
\end{tabular}

\end{center}

\end{table}

This experiment gives the first violation of the Bell inequality for massive
particles obtained using a complete set of correlation measurements. A previous
experiment \cite{proton} deduced a violation from an incomplete set of
measurements. Furthermore, our data uses the outcome of every shot of the
experiment, so the violation is obtained without the use of the fair-sampling
hypothesis. In principle, sampling or post-selection of the data allows
violation of Eq.~(\ref{belleq}) even by local hidden-variable theories, the
so-called ``detector loophole" \cite{detloop}. Our experiment closes the
detector loophole for the first time. However, the ``locality loophole"
\cite{locloop} remains open for our data; since the detection events on the two
ions occur within each other's lightcones, in principle the detections could
influence each other, leading to spurious correlations.\\ \\

\section{Encoding into a Decoherence-Free Subspace}

Information stored in a quantum system is liable to be lost through coupling of
the system to its environment, a process called decoherence
\cite{zurek,decohbook}. In particular, many proposed quantum computer
implementations decohere through a mechanism that couples each qubit to the
environment equally strongly \cite{forts1,forts2}. Encoding quantum information
into a decoherence-free subspace (DFS) \cite{lidar1,zanardi,duan} whose states
are invariant under coupling to the environment allows information storage even
in the presence of this type of decoherence. We have demonstrated the immunity
of a DFS of two atoms to environment-induced dephasing and
implemented a technique for encoding an arbitrary physical qubit state into the
DFS.\\ \\

The type of decoherence considered here is collective dephasing, in which each
qubit undergoes the transformation $\dk \rightarrow \dk, \uk \rightarrow
e^{i\zeta} \uk$ for $\zeta$ an unknown phase that is the same from qubit to qubit.
One can construct a DFS of two qubits that protects against collective dephasing
from the basis states $\udk, \duk$. Any superposition of these states is clearly
invariant under collective dephasing, and since there are two states, we can
encode one ``logical qubit" in this subspace of the two ``physical qubits". To
demonstrate our encoding technique, we first prepare a state of form

\begin{equation}
\label{init}
\dk(a\dk+b\uk) \hspace{0.5 cm} |a|^2+|b|^2=1 \hspace{0.5 cm} a, b\:\: \rm{complex}
\end{equation}

\noindent where one of the two physical qubits initially contains the quantum
information. In general, we can prepare this state by driving the carrier
transition of Eq. (\ref{careq}) on both ions simultaneously, once with
$\theta=\beta$, and again with $\theta=\beta$ and $\phi_1$ shifted by $\pi$. We
set $\phi_2=\alpha$ for both pulses. The final state has $a=\cos{2\beta},
b=e^{i\alpha}\sin{2\beta}$.\\ \\

We encode the state of Eq. (\ref{init}) into the DFS in two steps. First we
apply the inverse of the entangling gate, yielding
$a(\ddk-i\uuk) + b(\duk-i\udk)$. Then we drive the carrier with $\theta=\pi/4,
\phi_1=\pi/2, \phi_2=0$ to obtain $\dfs = a(\duk+i\udk) + b(\duk-i\udk)$. The information
stored in the physical qubit of ion 2 is now encoded in the logical qubit of the
DFS. In the experiment we take $|a| = |b|$, though our method permits
preparation and encoding of any state of the more general form.\\ \\

To read out the encoded information, we reverse the carrier pulse in the encoding
and apply the entangling gate to decode $\dfs$ into $\dk(a\dk+b\uk)$. After
decoding, we rotate ion 2 as in preparing the state of Eq. (\ref{init}) but with
the phase on ion 2 changed to $\alpha'$. We then measure the probability $P_2$
of finding both ions in $\dk$. $P_2$ varies sinusoidally with $\alpha-\alpha'$, and the
magnitude of oscillation is equal to the coherence, $C$, of
ion 2. Since we set $|a|=|b|=1/\sqrt{2}$, ideally $C=1$. Departures
from $C=1$ measure the effects of both decoherence and imperfect logic. We verified
that $C$ is independent of $\alpha$, thus showing that our encoding method works
even if we have no information about the phase of the input state.\\ \\

To study the effects of decoherence on the DFS-encoded state, we engineered a
collective dephasing environment using an off-resonant laser beam with a
randomly varying intensity. The beam induces shifts of $\nu_0$ that are
common to both ions through the AC Stark effect. The DFS state should resist the
dephasing effect of this environment. The coherence of ion 2 in the test state
$\dk(\dk+e^{i\phi}\uk)/\sqrt{2}$, however, should be sensitive to collective
dephasing. We measure the decay of the test state by simply turning off the
encoding and decoding sequences in the procedure used to measure the decay of
the DFS-encoded state. We applied the noise beam to the test state and encoded state
during a delay time of about 25 $\mu$s, as shown in Fig. 5. The coherence
without applied noise is about 0.69 for the test state and about 0.43 for the
encoded state; they depart from 1 because of imperfect logic gates and
detection. For white-noise intensity fluctuations of the Stark-shifting beam, we
expect $C$ to decay exponentially for the test state, as shown by the fit line.
We also fit the coherence of the DFS state to an exponential decay for
comparison. The decay rate of the test state is $0.18(1)\:\mu {\rm s}^{-1}$,
while the decay rate of the DFS state is $0.0035(50) \:\mu {\rm s}^{-1}$,
consistent with zero decay.\\ \\

\begin{figure}

\begin{center}

\epsfig{file=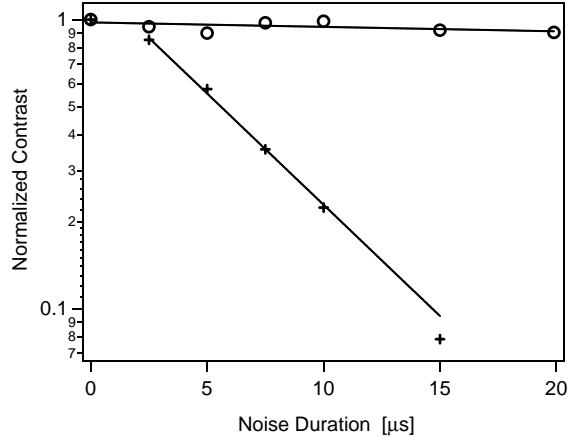, width=8cm, clip=, bbllx=80, bblly=350,
bburx=500, bbury=680}

\caption{Decay of DFS-encoded state (circles) and test state (crosses)
under engineered dephasing noise. The noise is applied for a fraction of the
delay time of about 25 $\mu$s between encoding and decoding. Coherence data are
normalized to their values for zero applied noise. The fit lines are exponential
decay curves. The DFS data is fit for comparison.}

\end{center}

\end{figure}

We have also measured the storage times of the encoded and test states under
ambient conditions in our laboratory. The data are shown in Fig. 6. Here we leave
a variable delay time between encoding and decoding to give the ambient noise
time to act. The decoherence of the test state is dominated by ambient
fluctuating magnetic fields whose frequencies lie primarily at 60 Hz and its
harmonics. These fields randomly shift $\omega_0$ through the Zeeman effect.
Since these fields are roughly uniform across the ion string, they induce
collective dephasing similar to that created by the engineered environment. We
empirically find the decay of both test and encoded states to be roughly
exponential, as indicated by the fit lines. The decay rate of the test state is
$7.9(1.5)\times 10^{-3} \:\mu {\rm s}^{-1}$, while the decay rate of the DFS
state is $2.2(0.3)\times 10^{-3} \:\mu {\rm s}^{-1}$. The long lifetime of the
DFS state shows that collective dephasing from magnetic field noise is the major
ambient source of decoherence for the test state. The loss of coherence of the
encoded state is consistent with degradation of the decoding pulses
\cite{molmer,bible} due to heating of the ion motional state over the delay time
\cite{quentin}.\\ \\

\begin{figure}

\begin{center}

\epsfig{file=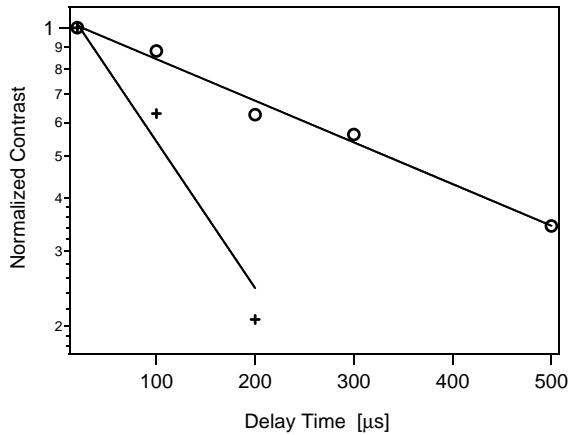, width=8cm, clip=, bbllx=80, bblly=340,
bburx=500, bbury=660}

\caption{Decay of DFS-encoded state (circles) and test state (crosses)
under ambient decoherence. We vary the delay time between encoding and decoding
to give the ambient noise a variable time to act. Coherence data are normalized to
their values for zero delay time. The fit lines are exponential decay curves
for purposes of comparison and are not theoretical predictions.}

\end{center}

\end{figure}

\section{Conclusion and Outlook}

We have demonstrated four-particle entanglement, violation of Bell's inequality,
and encoding into a decoherence-free subspace using a small quantum register of
trapped ions. While these experiments are important steps on the road to quantum
computation, the major challenge remains: scaling up the techniques developed
for a few ions to build a large-scale quantum computer with thousands or
millions of qubits. Fortunately, we have a plausible roadmap to our goal, the
``quantum CCD" model of ion-trap quantum computing (Fig. 7). We envision a large
ion trap with many independent DC electrodes, which could be used to move ions
between storage regions and ``accumulator" regions in which quantum logic
operations take place. Such a trap could hold thousands of ions, but avoids the
slow gate speeds and detection inefficiency associated with large ion crystals
by allowing us to work with only a few qubits at a time. Though the ions would
most likely heat up while being moved around, the addition of another ion
species would enable us to sympathetically cool the ions before performing logic
\cite{symp}. The DFS encoding demonstrated above would permit us to move qubits
around without incurring dephasing from the spatial variation of magnetic fields
across the device. In pursuit of the ``quantum CCD" goal, we have built a trap
supporting a double-well potential with spacing about 1 mm and plan to explore
ways of maintaining quantum coherence and entanglement over these large
distances.\\ \\

\begin{figure}[top]

\begin{center}

\epsfig{file=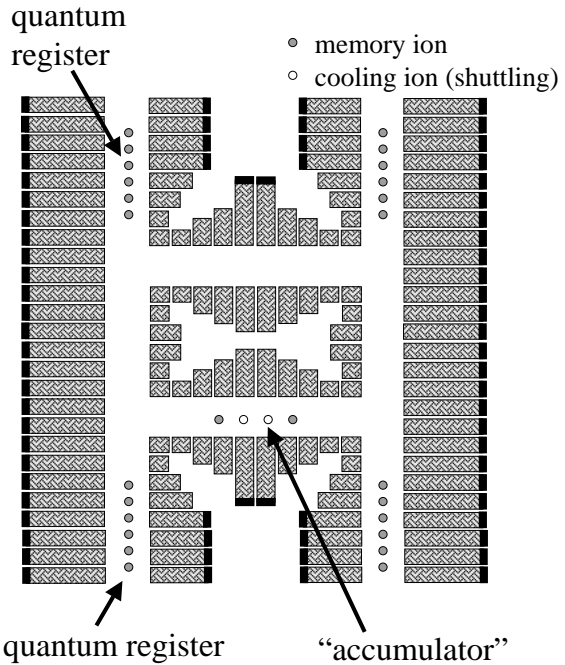, width=8cm, clip=, bbllx=100, bblly=130,
bburx=540, bbury=640}

\caption{Schematic of the proposed ``quantum CCD" architecture for a large-scale
ion-trap quantum computer. The ``logic ions", which encode the quantum
information, are held in a storage region. To perform a logic operation, we
shuttle the relevant logic ions into the ``accumulator" region, where they
interact with lasers. Auxiliary ``cooling ions" are provided for sympathetic
cooling of the logic ions, which will almost certainly heat up during transfer to the
accumulator.}

\end{center}

\end{figure}

\acknowledgments{This research was supported by the NSA, ARDA, ARO, and ONR.
This paper is a contribution of the National Institute of Standards and
Technology and is not subject to U.S. copyright.}

\end{document}